\documentstyle[12pt,epsf]{article}
\begin{document}

\begin{center}
{\large \bf Residual resistivity due to wedge disclination dipoles
in metals with rotational plasticity\\} \vskip 1.5em
{\large S.E.Krasavin \footnote{Corresponding author. \\
\hspace*{0.5cm}{\it E-mail address:} krasavin@thsun1.jinr.ru (S.E. Krasavin)}\\
\vskip .5em
\small\it Joint Institute for Nuclear Research\\
Bogoliubov Laboratory of Theoretical Physics\\
141980 Dubna, Moscow region, Russia\\}
\vskip 1em
\end{center}

\begin{abstract}
The residual resistivity $\rho $ in metals caused by wedge
disclination dipoles is studied in the framework of the Drude
formula. It is shown that $\rho\sim L^{-p}$ with $p=3$ for biaxial
and $p=2$ for uniaxial dipoles ($L$ is a size of  dipole arm).
\end{abstract}



\newpage

The effect of dislocations on electric transport in metals has
been  studied for many decades [1,2,3]. Dislocations serve as
effective scattering centers for conducting electrons primarily
due to their elastic strain fields. This scattering is essential
in the region of the residual resistivity at low temperatures when
all other scattering mechanisms are suppressed. However, the
problem of disclination-induced charge scattering is not yet
investigated in details, despite the fact that these linear
defects can play an important role in nanocrystalline [4] and
highly deformed metals [5]. Such defects, combined in dipole
configurations, have been proposed as primary carries of the
rotational plastic deformation in granular materials (see e.g. [6]
and references therein) and observed recently in nanocrystalline
Fe [7] using the high-resolution transmission electron microscopy.
For metallic glasses the concept of disclinations has been worked
out in [8] and much earlier for complex alloys in [9].

Theoretically, for the first time, the behaviour of the residual
resistivity as a function of the density of defects in simple
metals caused by {\it isolated} wedge disclinations has been
studied in [10]. The analysis has been carried out with the
assumption that there exist two mechanisms of scattering: due to
deformation fields of wedge disclinations and Aharonov-Bohm-like
scattering generated by topological nature of disclinations [11].
The deviation from the linear law of the disclination-induced
residual resistivity on the concentration of the defects was
found.

In this Letter we study the behaviour of the residual resistivity
in metals containing wedge disclination dipoles (WDD). Our goal is
to find the dependence of the residual resistivity on the value of
the dipole arm $L$.  In fact, keeping in mind the models where
disclinations are settled in the triple junctions of inter grain
boundaries [12,13] or form borders of misorientation band area
[14,15], we study how residual resistivity depends on a grain size
or  a width of misorientation band.

 On the other hand, it was found (see e.g. [16,17]) that strain fields
caused by WDD are the same as for a finite wall of edge
dislocations at large distances from the wall. Hence, the obtained
here results can be considered in application to the materials
containing dislocation arrays and small-angle grain boundaries.
 In our picture the dipoles in equilibrium with a mean dipole arm $L$ and
 strength $\pm\omega$ are placed in $xy$-plane (disclination lines
 are oriented along the  $z$-axis). Notice that a disordered distribution of
disclination lines only modify the absolute value of a electron
mean free path in our calculations. The axes of the rotations can
be shifted relative to their lines by arbitrary distances $l_{1}$
and $l_{2}$. When $l_1-l_2=L$ or $l_1=-l_2$  one gets the uniaxial
and symmetrical uniaxial WDD, respectively. In the case when
$l_1\ne l_2\ne 0$, we have biaxial WDD with shifted axes of
rotation (see, e.g., [17,18]).

The effective perturbation energy of electron due to the WDD
deformations $E_{AB}$ is [18]
$$
U({x,y})=-G_{d}SpE_{AB}(r)=
$$
\begin{equation}
 -\frac{G_{d}(1-2\sigma  )\omega}{(1-\sigma)4\pi}
\left(\ln\frac{(x+L/2)^2+y^2}{(x-L/2)^2+y^2}-l_{1}
\frac{x+L/2}{(x+L/2)^2+y^2}+l_{2}\frac{x-L/2}{(x-L/2)^2+y^2}\right),
\end{equation}
where $G_{d}$ is the deformation-potential constant, $\sigma$ is
the Poisson ratio. For simplicity, in Eq.(1) we have considered
only isotropic component of the deformation-potential constant,
which is related to the Fermi energy as $(2/3)E_{F}$ [2]. In this
context, in further calculations we use the typical meaning of
$G_{d}=3.7$eV.
 It is seen from the Eq.(1) that the WDD strain fields are
located in $xy$-plane. It means that only normal to disclination
line component of electron wave vector ${\bf k}_\perp $ are
involved in scattering process. As a result, the problem reduces
to the two-dimensional scattering where the matrix element which
determines the transition of electron from Fermi state with wave
vector ${\bf k}_{F}={\bf k}_{\perp} +{\bf k_z}$ to state ${\bf
k}^{'}$ can be written as [2,18]
\begin{equation}
{\left\langle{\bf k}_F|U(\rho,\phi)\right|{\bf
k'}\rangle}=\frac{1}{S}\int d^2\rho\exp[i(k_{F}-k^{'})\rho
\cos(\phi -\alpha )]U(\rho,\phi).
\end{equation}
Here, $S$ is the projected area, $U(\rho,\phi)$ is perturbation
energy given by Eq.(1) in polar coordinates $(\rho,\phi)$,
$\alpha$ is the angle between ${\bf q}={\bf k}_{F}-{\bf k}^{'}$
and $x$-axis.

 Using Eqs.(1) and (2) with the general formula for the two-dimensional mean
 free path
\begin{equation}
l^{-1}=\frac{n_dk_F S^2}{2\pi\hbar ^2v_F
^2}\int_{0}^{2\pi}\left(1-\cos\theta\right)\overline{\left|\langle{\bf
k}_F|U(x,y)\right|{\bf k'}\rangle|^2}d\theta,
\end{equation}
after integration over scattering angle $\theta $, we find the
explicit expression for the mean free path as
\begin{eqnarray}
l^{-1}=\frac{B^2L^2n_{d}\pi ^2}{4k_{F}\hbar
^2v_{F}^2}\!\!\!\!\!\!\!\!\! &&\left\{
z^2\left(\frac{1}{2}+J^2_{0}(k_{F}L)\right)+\left(8-\frac{z(z+8)}{2}\right)(J^2_{0}(k_{F}L)
\right.\nonumber\\
&& +\left. J^2_{1}(k_{F}L))
-\frac{8}{k_{F}L}J_{0}(k_{F}L)J_{1}(k_{F}L)\right\},
\end{eqnarray}
where $z=2(l_{1}-l_{2})/L$, $B=G_{d}\omega (1-2\sigma)/(1-\sigma
)2\pi $, $v_{F}$ is the Fermi velocity, $J_{n}(t)$ are the Bessel
functions. In Eqs.(3) and (4) $n_{d}$ is the areal density of the
dipoles, and the bar in Eq.(3) denotes the averaging over $\alpha
$.

Evidently, $n_{d}$  is a function of the dipole arm $L$. To
determine the relation between $n_{d}$ and  $L$, notice, that for
two dimensional elastically isotropic medium $n_{d}$ is inversely
proportional to the square of the  mean distance $d$ between
dipoles. In the framework of the dislocation-disclination model of
misorientation band [15] the dependence of $d$ on $L$ at the state
of equilibrium can be found from the relation
\begin{equation}
qb=\omega d\ln\left(\frac{L^2}{d^2}+1\right),
\end{equation}
where $b$ is the absolute value of a misorientation band Burgers
vector, $q\geq 1$ is a dimensionless parameter which account the
presence of "statistically-stored" dislocations. For the case when
$d>L$ we have $d\approx \omega L^2/qb$, and $n_{d}\approx
1/d^2$=$\left(qb/\omega L^2\right)^2$

 Our analysis shows that for the mean free path given by Eq.(4)
 the condition $k_{F}l\gg 1$ is valid and the classical
Drude formula to estimate the residual resistivity can be applied
\begin{equation}
\label{eq1} \rho=\left(\frac{m v_{F}}{ne^{2}}\right) l^{-1},
\end{equation}
where $m$ and $n$ denote mass and electron density, respectively,
$e$ is the electron charge.

Thus, the $L$-dependence of the residual resistivity $\rho $ can
be defined numerically on the basis of the Eqs.(4)-(6). The
results of the calculations are shown in Fig.1 for all types of
WDD with strength $\omega =36^{\circ}$.
\begin{figure}[h]
\epsfxsize=9cm \centerline{\hspace{4mm} \epsffile{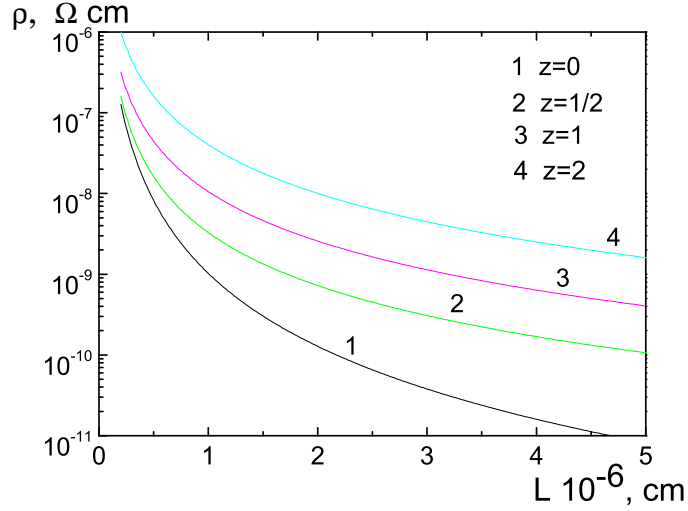}}
\vspace{3mm} \caption{ The residual resistivity as a function of
the dipole arm of size $L$ for symmetrical biaxial disclination
dipole ($z=0$); biaxial dipole with shifted axes of rotation
($z=1/2$, $z=1$); uniaxial dipole ($z=2$). The curves have been
plotted according to Eq.(4) and Eq.(6) with the set of the
parameters: $B=0.1$eV, $v_{F}\approx1.2\times10^8$cm c$^{-1}$,
$n=5\times10^{22}$cm$^{-3}$, $m=0.5\times10^{6}$eV } \label{fig1}
\end{figure}

 As is seen from the plot the least contribution to
$\rho $ is caused by WDD with $z=0$, (i.e. $l_1=l_2$ that
corresponds to the symmetrical  biaxial dipole), and $\rho $
increases with $z$ increasing. For $z=2$ (uniaxial WDD) the
contribution to $\rho $ is the largest.  This noticeable increase
of $\rho (z=2)$ relative to $\rho (z=0)$ is due to  the specific
nature of the uniaxial WDD deformation fields. Uniaxial WDD can be
simulated by a finite wall of edge dislocations complemented by
two additional edge dislocations at both ends of the wall [17].
These two dislocations are represented in Eq.(1) by the second and
third terms. Obviously, the residual resistivity due to a uniaxial
WDD has a larger value due to the presence of this dislocation
part which is absent for biaxial WDD. It should be noted that the
functional $L$-dependence of $\rho $ is different for biaxial and
uniaxial dipoles. Indeed, $l^{-1}\sim Ln_{d}$  for Eq.(4) in the
limit $k_{F}l\gg 1$ when $z=0$. Taking into account the relation
$n_{d}\sim L^{-4}$, we find for biaxial dipole $\rho(z=0)\sim
l^{-1}\sim L^{-3}$. In the case of the uniaxial dipoles
$l^{-1}\sim L^2n_{d}$, and, hence $\rho (z=2)\sim l^{-1}\sim
L^{-2}$.

The  important result of our consideration here is that the
residual resistivity  increases when $L$ (or,equivalently,
granular size) decreases. It is easily understood, because in our
approach the $L$-dependence of $n_{d}$ has been considered
correctly  in the framework of the misorientation band model [15].
In [19] the increase of $\rho $  with grain-size decreasing has
been found experimentally for nanocrystalline Pd. These results
are in qualitative agreement with our calculations.

Fig.2 demonstrates the $n_{d}$-dependence of $\rho$ for uniaxial
WDD with different strengths of defects $\omega $. This dependence
is nonlinear ($\rho(z=2)\sim n_{d}^{1/2}$) as one can conclude
from the previous reasonings. The nonlinear dependence of $\rho $
has been found in [10] for isolated wedge disclinations as well.
Similar result  should be expected for edge dislocation walls as
we have discussed in the beginning of this paper.
\begin{figure}[h]
\epsfxsize=9cm \centerline{\hspace{4mm} \epsffile{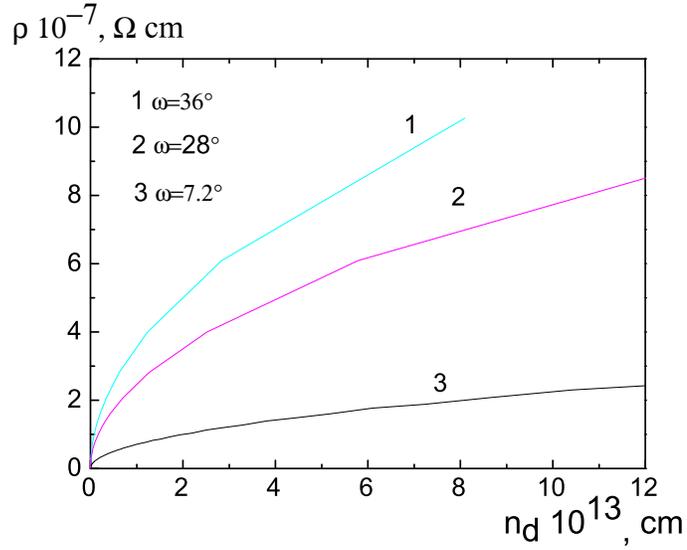}}
\vspace{3mm} \caption{ The residual resistivity as a function of
the arial density of uniaxial dipoles $n_{d}$ at different defect
strengths $\omega $. } \label{fig2}
\end{figure}
Let us note that linear $n_{d}$-dependence of $\rho $ had been
observed only for isolated dislocations (See [3], and references
therein). In addition, one can see from  Fig.2 that $\rho $
increases substantially with increasing $\omega $ reaching quite
large values. For example, for the curve number one $\rho\approx
6\times 10^{-7}$ $\Omega$ cm when $n_{d}\approx 3\times 10^{13}$
cm$^{-2}$ (that correspond to the dipole arm $L$
 equal to few nanometers).

In conclusion, we would like to mention, that the resistivity due
to oriented in some direction disclination dipoles should be
anisotropic (as in the case of dislocations [2]). For example, for
edge dislocations with glide direction along the $x$-axis, the
ratio $\rho _{x}/\rho_{y}$ has been found to be equal to
$\frac{1}{3}$ [20,21]. Calculations of $\rho $ in different plane
directions for disclination dipoles (dislocation walls) will be
performed in the near future.

\newpage

\centerline{Figure Captions}
\vskip 1cm

\noindent Fig.1.The residual resistivity as a function of the
dipole arm of size $L$ for symmetrical biaxial disclination dipole
($z=0$); biaxial dipole with shifted axes of rotation ($z=1/2$,
$z=1$); uniaxial dipole ($z=2$). The curves have been plotted
according to Eq.(4) and Eq.(6) with the set of the parameters:
$B=0.1$eV, $v_{F}\approx1.2\times10^8$cm c$^{-1}$,
$n=5\times10^{22}$cm$^{-3}$, $m=0.5\times10^{6}$eV  .
\vspace{1.5cm}

\noindent Fig.2.The residual resistivity as a function of the
arial density of uniaxial dipoles $n_{d}$ at different defect
strengths $\omega $ .

\end{document}